\documentclass[reprint,prl,twocolumn,showpacs,superscriptaddress,aps]{revtex4-1}
\usepackage[utf8]{inputenc}

\usepackage{graphicx}
\DeclareGraphicsExtensions{.png,.jpg,.eps}

\usepackage{xcolor}
\usepackage{amsmath}
\usepackage{caption}
\usepackage{subcaption}
\usepackage{float}

\begin{document}

\title{Absence of ferromagnetism in VSe$_2$ caused by its charge density wave phase}

\author{Adolfo O. Fumega}
  \email{adolfo.otero.fumega@usc.es}
\author{Victor Pardo}
  \email{victor.pardo@usc.es}
\affiliation{Departamento de F\'{i}sica Aplicada,
  Universidade de Santiago de Compostela, E-15782 Campus Sur s/n,
  Santiago de Compostela, Spain}
\affiliation{Instituto de Investigaci\'{o}ns Tecnol\'{o}xicas,
  Universidade de Santiago de Compostela, E-15782 Campus Sur s/n,
  Santiago de Compostela, Spain}


\begin{abstract}

In this study we present a detailed \emph{ab initio} analysis of the magnetic properties of VSe$_2$. Ab initio calculations in the so-called 1T structure yield a ferromagnetic phase as most stable, with a magnetic moment of about 0.6 $\mu_B$/V. According to our calculations this ferromagnetic state is on the verge of instability. 
We have modeled ab initio the charge density wave state reported in the literature. This introduces a periodic lattice distortion leading to a supercell with periodicity 4$a$ $\times$ 4$a$ $\times$ 3$c$ (2$a$ $\times$ 2$a$ for the monolayer) in which we have fully relaxed the atomic positions. 
We demonstrate that this structural rearrangement causes a strong reduction in the density of states at the Fermi level and the ground state of the system becomes non-magnetic for the bulk. 
In the monolayer limit the rearrangement induces a Peierls distortion causing an energy gap opening at the Fermi level and the quenching of ferromagnetism.

\end{abstract}

\maketitle

Since the discovery of graphene\cite{Graphene} there has been an enormous scientific effort in the search and characterization of purely two-dimensional (2D) materials that could show new physical properties and lead to potential new applications. In particular, the family of layered transition metal dichalcogenoides (TMDs) has been widely studied over the past years\cite{manzeli_2d_2017,CHOI2017116,C7TC01088E,XIA20171}.

Very recently, ferromagnetism has been observed in a purely 2D material\cite{huang_layer-dependent_2017}, and since then, the field of ferromagnetic 2D materials has gained momentum with the appearance of various families of layered, van der Waals bonded materials that remain ferromagnetic when exfoliated to the ultrathin limit \cite{gong_discovery_2017,VS2_FM}.

\begin{figure}[!h]
\begin{center}
\includegraphics[width=0.40\columnwidth]{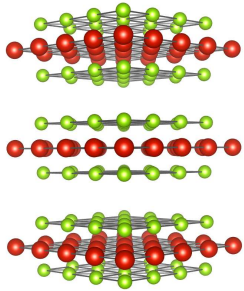}
\end{center}
\caption{(Color online.) Representation of the 1T-structure as experimentally described for VSe$_2$. V atoms are in red and Se atoms in green. Observe the layered structure, layers bond weakly via van der Waals forces.}
\label{struct}
\end{figure}

Vanadium diselenide (VSe$_2$) is a TMD that crystallizes naturally in a trigonal (T) phase. This consists of V atoms 6-fold coordinated by Se atoms forming layers in the (0001) direction 
(see Fig. \ref{struct}). 
Van der Waals forces are responsible for the weak bonding between adjacent layers. 
As many other materials of that sort, VSe$_2$ is relatively easy to exfoliate\cite{xu_ultrathin_2013} or grow\cite{bonilla_strong_2018,nano_monolayer} till the single and few-layer limit.

It has been reported that various TMDs show a charge density wave (CDW) phase at low temperatures\cite{TiSe2_CDW, TMD_CDW}. In the case of VSe$_2$ this phase appears below 110 K\cite{VSe2_CDW1}. 
That critical temperature increases when going to the 2D scenario\cite{xu_ultrathin_2013} and also when pressure is applied\cite{FRIEND1978169}.  
Signatures of the transition are seen in transport properties such as resistivity\cite{van_bruggen_magnetic_1976} or thermopower\cite{yadav_electronic_2010}.  
A symmetry breaking of the space group has been demonstrated to occur if a CDW transition exits\cite{johannes_fermi_2008}. 
VSe$_2$ belongs to the P-3m1 space group in the normal state (NS), whereas a commensurate 4$a$ $\times$ 4$a$ $\times$ 3$c$ supercell arises in the CDW state\cite{van_bruggen_magnetic_1976, 0022-3719-10-14-013,PhysRevLett.109.086401} that occurs at low temperatures. 
This kind of periodic lattice distortion introduces extra reflections in the diffraction patterns\cite{williams_charge_1976}.
Strain engineering\cite{zhang_strain_2017} or atom intercalation\cite{PhysRevB.59.7751} can be used to change the modulation of the CDW supercell.

The T-phase of VSe$_2$, the one that we are going to study, presents a metallic behaviour in both the NS and the CDW state\cite{BAYARD1976325} for the bulk structure. 
However, when going to the 2D limit an energy gap opening at the Fermi level in the whole Brillouin zone is reported to occur in the CDW state\cite{nano_monolayer}. 
Experiments report a paramagnetic behaviour of bulk VSe$_2$\cite{van_bruggen_magnetic_1976, BAYARD1976325,barua_signatures_2017,cao_defect_2017}.
When going to the monolayer case the magnetic behavior of this system is not totally clear. Some articles state a non-magnetic situation\cite{nano_monolayer}, while others claim ferromagnetism (FM) arises\cite{xu_ultrathin_2013,bonilla_strong_2018}.  
However, a huge discrepancy in the saturation magnetization is seen between a VSe$_2$ monolayer grown on top of MoS$_2$ ($\sim 15$ $\mu_{B}$ per V atom)\cite{bonilla_strong_2018} and the exfoliated few-layer-thick VSe$_2$  films ($\sim 0.3 \times 10^{-3}$ $\mu_{B}$ per V atom)\cite{xu_ultrathin_2013}.   
Previous density functional theory (DFT) calculations of bulk and monolayer VSe$_2$ show that a FM phase is the most stable one\cite{li_versatile_2014, ma_evidence_2012}, but with values of the magnetic moment in strong disagreement with those experimentally obtained, yet consistent with our own set of calculations (see below). Previous DFT-based ab initio calculations show that perturbations to the system, such as strain, are able to destroy magnetism\cite{ma_evidence_2012}.

In this work, we will try to understand the magnetic properties of VSe$_2$ from an \emph{ab initio} perspective.
For that sake, we have performed \emph{ab initio} electronic structure calculations based on DFT\cite{HK,KS} 
using an all-electron full potential code ({\sc wien2k}\cite{WIEN2k}) on VSe$_2$.
The exchange-correlation term used for the bulk structure was the generalized gradient
approximation (GGA) in the Perdew-Burke-Ernzerhof\cite{PBE} scheme for structural optimizations and to compute the energetics. 
The LDA+U method was used for the 2D case\cite{LDAU}. This allows to find a good description of the electronic structure of the CDW in the single-layer limit. 
These calculations were performed with a converged k-mesh and a value of  R$_{mt}$K$_{max}$= 7.0.
The R$_{mt}$ value used was 2.12 in a.u. for both V and Se. Structural data was taken from Ref. \onlinecite{VSe2_lattice}.
Transport properties have been calculated using the BoltzTrap2 code\cite{Boltztrap2}. This solves Boltzmann transport equation from first-principles calculations within the constant scattering time approximation. A denser k-mesh was used for this task.

\begin{figure}[!h]
\begin{center}
\includegraphics[width=\columnwidth]{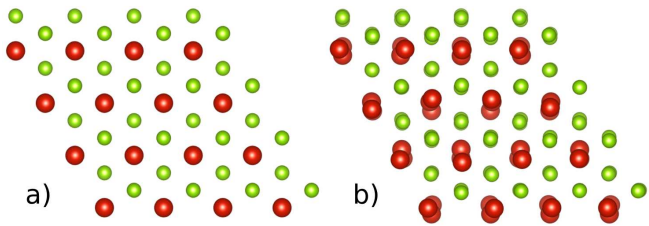}
\caption{(Color online.) Top view of VSe$_2$ bulk structures. V atoms are represented as big red spheres and Se atoms are shown in green. 
a) NS structure. It can be reduced to a three-atom unit cell that belongs to the P-3m1 space group. DFT predicts a FM ground state for it. 
b) CDW structure. A modulated 4$a$ $\times$ 4$a$ $\times$ 3$c$ supercell appears in the CDW state. FM is absent in this situation.}
\label{structures}
\end{center}
\end{figure}

\begin{figure}[!h]
\begin{center}
\includegraphics[width=\columnwidth]{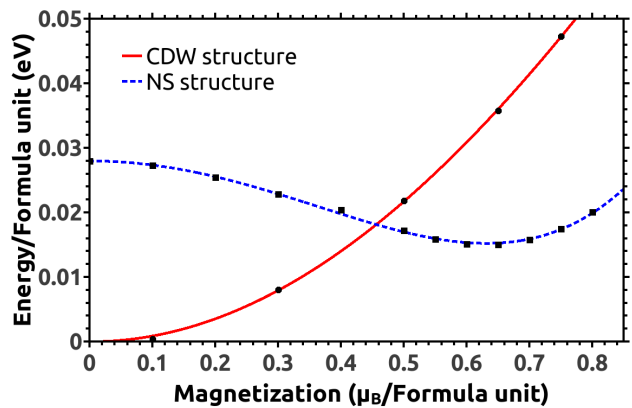}
\end{center}
\caption{(Color online.) Energy as a function of the magnetization for the bulk structures. Computed points in black. The blue dashed line corresponds to the NS structure. 
It presents a minimum at around $0.6$ $\mu_B$ per V atom. 
The red line corresponds to the CDW structure. It can be seen that the minimum-energy configuration is non-magnetic and also the ground state of VSe$_2$.}
\label{E_vs_M_bulk}
\end{figure}

The NS bulk structure in the P-3m1 space group  can be seen in Fig. \ref{structures}a. The top view shows the hexagonal symmetry. The calculations yield a FM ground state with a total moment of $0.6$ $\mu_B$ per V atom. The total energy as a function of the magnetization can be seen as the blue dashed line in  Fig. \ref{E_vs_M_bulk}, with a somewhat broad minimum around that value of the magnetization.

To understand the origin and characteristics of this ferromagnetic ground state, we can start thinking of VSe$_2$ as an itinerant electron system and use the phenomenological Stoner theory to determine if the FM phase is stable\cite{Stoner_crit} or not. The Stoner criterion makes a comparison between the energy gained by the system via a spin splitting compared to the kinetic energy cost produced by displacing minority-spin electrons into a higher-energy majority-spin band. Only when the overall energy gets reduced an itinerant electron system like this can become spontaneously magnetic. This is usually formulated in the following way: 

\begin{equation}\label{stoner_c}
  \text{Stoner criterion}\begin{cases}
    \text{FM, } \text{if $I\cdot DOS(E_F)>1$}.\\
    \text{Non-magnetic, } \text{otherwise}.
  \end{cases}
\end{equation}

where $I$ is the exchange energy between the Bloch d-band electrons, the so called Stoner parameter. 
It can be obtained from the energy (E) \emph{vs} magnetization (M) fitting curve, $E=(1-I DOS(E_F))/DOS(E_F) M^2 +k M^4$, plot in Fig. \ref{E_vs_M_bulk}, where $k$ is a fitting parameter independent of $I$. This procedure is detailed in Ref. \cite{moriya1985spin}
The density of states for the non-magnetic solution at the Fermi level ($DOS(E_F)$) can also be computed \emph{ab initio}.

\begin{figure}[!h]
\begin{center}
\includegraphics[width=\columnwidth]{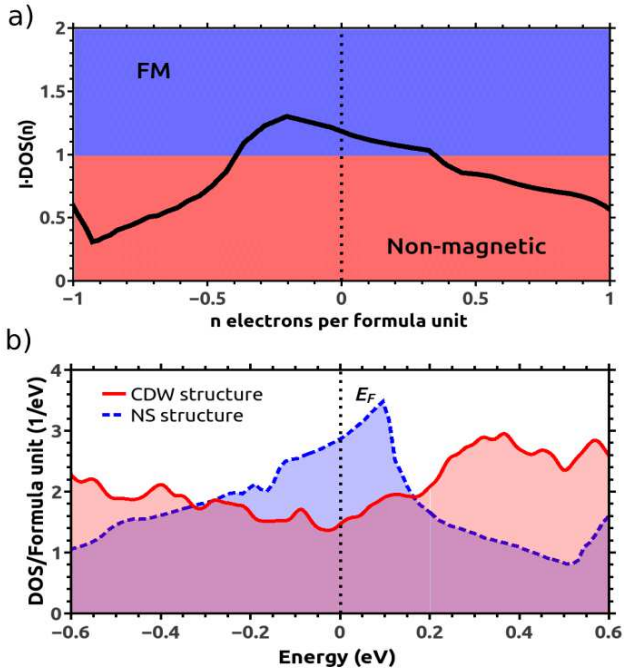}
\end{center}
\caption{(Color online.) a) Stoner criterion for the NS structure as a function of the number of electrons introduced per formula unit. When $I\cdot DOS>1$ the system is said to be 
FM. VSe$_2$ in the NS is FM but somewhat close to a non-magnetic state being stable. 
b) DOS around the Fermi level for the bulk structures. DOS of the CDW is shown in red, while the blue dashed represents the DOS of the NS. 
There is a clear reduction of the DOS when the CDW is present. This contraction leads to the FM phase being unstable in the CDW state.}
\label{stoner_111}
\end{figure}

In Fig. \ref{stoner_111} a) we show how the Stoner criterion gets satisfied or not as a function of the number of electrons introduced per formula unit in the system by plotting the product of I$\cdot$DOS(E$_F$). 
The Fermi level corresponds to $n=0$ in the plot and moving to the right or the left implies hole or electron doping, respectively. The carrier concentration was calculated using a rigid band approximation by integrating the total density of states of the non-magnetic calculation. Applying (\ref{stoner_c}) we observe that the NS structure is inside the FM part of the phase diagram since the product of the Stoner parameter times the density of states at the Fermi level is larger than 1, but not by much. Let us recall that such product is about 2.5-3 for Fe, Ni, Co, the simplest itinerant ferromagnets\cite{stoner_atomicnumber}. Any perturbation to this system that could cause a small reduction in the DOS at the Fermi level would lead to a non-magnetic situation to become stable. For example, previous \emph{ab initio} studies have shown that a reduction of the FM moment can be achieved using strain engineering\cite{ma_evidence_2012} in this system, so that relatively small values of strain could make ferromagnetism disappear in VSe$_2$. The use of the Stoner criterion helps us understand in an oversimplified way the reason behind this set of results.

As mentioned above, VSe$_2$ presents a CDW state at low temperatures. It is known that a CDW state leads to gap openings around the Fermi level for particular values of the lattice momentum \cite{terashima_charge-density_2003}. Thus, the introduction of a periodic lattice distortion of that sort could have an effect in the calculated DOS at the Fermi level and hence in the magnetic properties of this itinerant electron system. In order to take this into account in our calculations, we have computed a 4$a$ $\times$ 4$a$ $\times$ 3$c$ supercell. The periodicity of the distortion was chosen from the experimental evidences that exist of the nature of the CDW in this system at low temperatures\cite{van_bruggen_magnetic_1976}.
For such an enlarged unit cell, we have optimized all the atomic positions obtaining the structure depicted in Fig. \ref{structures} panel b). One can see that the short-range hexagonal symmetry, in particular for the V sublattice, is lost, with the nearest neighbouring V-V distance becoming largely distorted.
This optimized CDW structure is $28$ $meV$  per formula unit more stable than the NS structure, a sizable value. 
Our result in the CDW structure is that the DOS at the Fermi 
level is vastly reduced compared to the NS structure. This can be seen in Fig. \ref{stoner_111} b) in which the CDW $DOS(E_F)$ (red line) is presented, being almost half that of the NS (blue dashed) one. Again, one can try to understand this based on a phenomenological Stoner-type description. The vanishing magnetization can be related to the large reduction in the DOS at the Fermi level.
This reduction of the DOS causes the FM moment to be quenched. Figure \ref{E_vs_M_bulk} shows the plot of the total energy as a function of the magnetization for the CDW state (red line), where the minimum is at zero moment, as opposed to the non-vanishing magnetization that is the ground state for the NS structure.

\begin{figure}[!h]
\begin{center}
\includegraphics[width=\columnwidth]{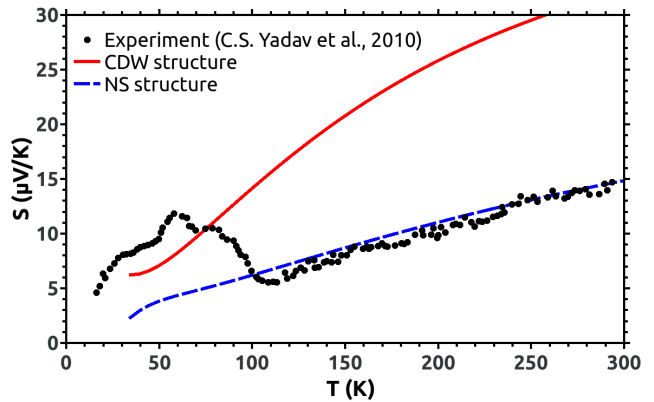}
\end{center}
\caption{(Color online.) Thermopower as a function of temperature for bulk VSe$_2$. 
The black points show the experimental measurements from Ref. \onlinecite{yadav_electronic_2010}. 
The red line shows the calculated thermopower for the 4$a$ $\times$ 4$a$ $\times$ 3$c$ supercell. 
It fits the experimental data at low temperatures, when the CDW is present. 
The blue dashed line shows the calculated thermopower for the P-3m1 cell. 
It fits the experimental data at high temperatures, when the NS is present.}
\label{seebeck}
\end{figure}

In order to give further evidences that the relaxed structure we have obtained can model reasonably well the CDW state found experimentally with a periodic lattice distortion in the form of a 4$a$ $\times$ 4$a$ $\times$ 3$c$ supercell, we have computed the thermopower of both the NS and the CDW structures and compared it to the experimental literature. 
Figure \ref{seebeck} shows the calculated thermopower as a function of temperature for the CDW (red line) and the NS structure (blue dashed line), and compared it with 
 the experimental data from Ref. \onlinecite{yadav_electronic_2010} (black points). Experimentally, a significant drop in the thermopower is observed at the transition from the CDW at low temperatures to the NS above 100 K. Our results show that, at any temperature, the Seebeck coefficient is higher for the CDW structure than for the NS one. 
This can be explained considering that the CDW phase opens pseudo-gaps in the Fermi surface and hence 
thermopower increases when gaps are opened around the Fermi level. If a crossover from the CDW at low temperatures to the NS structure about 100 K is to be expected, then we can see that our calculations help understand the behavior found in the thermopower reasonably well.

The CDW phase is experimentally observed in VSe$_2$ down to the monolayer limit, in that case even with a higher critical temperature\cite{xu_ultrathin_2013, bonilla_strong_2018}. 
This strongly suggests that the periodic lattice distortion associated to it needs to be considered when studying its magnetic properties. 
For that reason, we have carried out calculations for the monolayer as well. 
However, in the monolayer situation the hoping parameter t between layers vanishes and hence the parameter U$/$t, U being the on-site Coulomb repulsion, increases. 
In order to take into account the effect of correlations, we have used the LDA+U method. 
We have fully relaxed a 2$a$ $\times$ 2$a$ supercell for different values of U. 
The supercell that modulates the CDW is not totally clear for the monolayer\cite{nano_monolayer}. 
Our solution will be an energy minimum that will show the incopatibility between intrinsic FM and a CDW state. We do not claim we have found the overall ground state of the system since we have not explored all possible supercells, but we have found a local ground state for the 2$a$ $\times$ 2$a$ supercell that is in agreement with experiments and helps understanding the physics of the compound. Our calculations confirm that in order to do this it is mandatory to consider the appropriate lattice distortion associated to the CDW state, both in the bulk and in the monolayer.

\begin{figure}[!h]
\begin{center}
\includegraphics[width=\columnwidth]{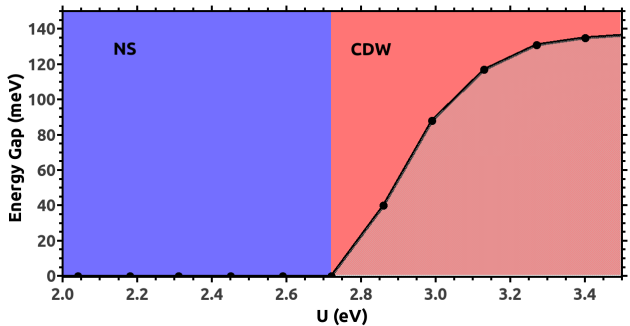}
\end{center}
\caption{(Color online.) Mapping on U for the 2$a$ $\times$ 2$a$ monolayer supercell. The structure was fully optimized for each value of U. 
At low values of U the metallic FM NS structure is the most stable. However for U$>2.7$ eV a non-magnetic CDW is formed and an energy gap arises.}
\label{gapU}
\end{figure}

\begin{figure}[!h]
\begin{center}
\includegraphics[width=\columnwidth]{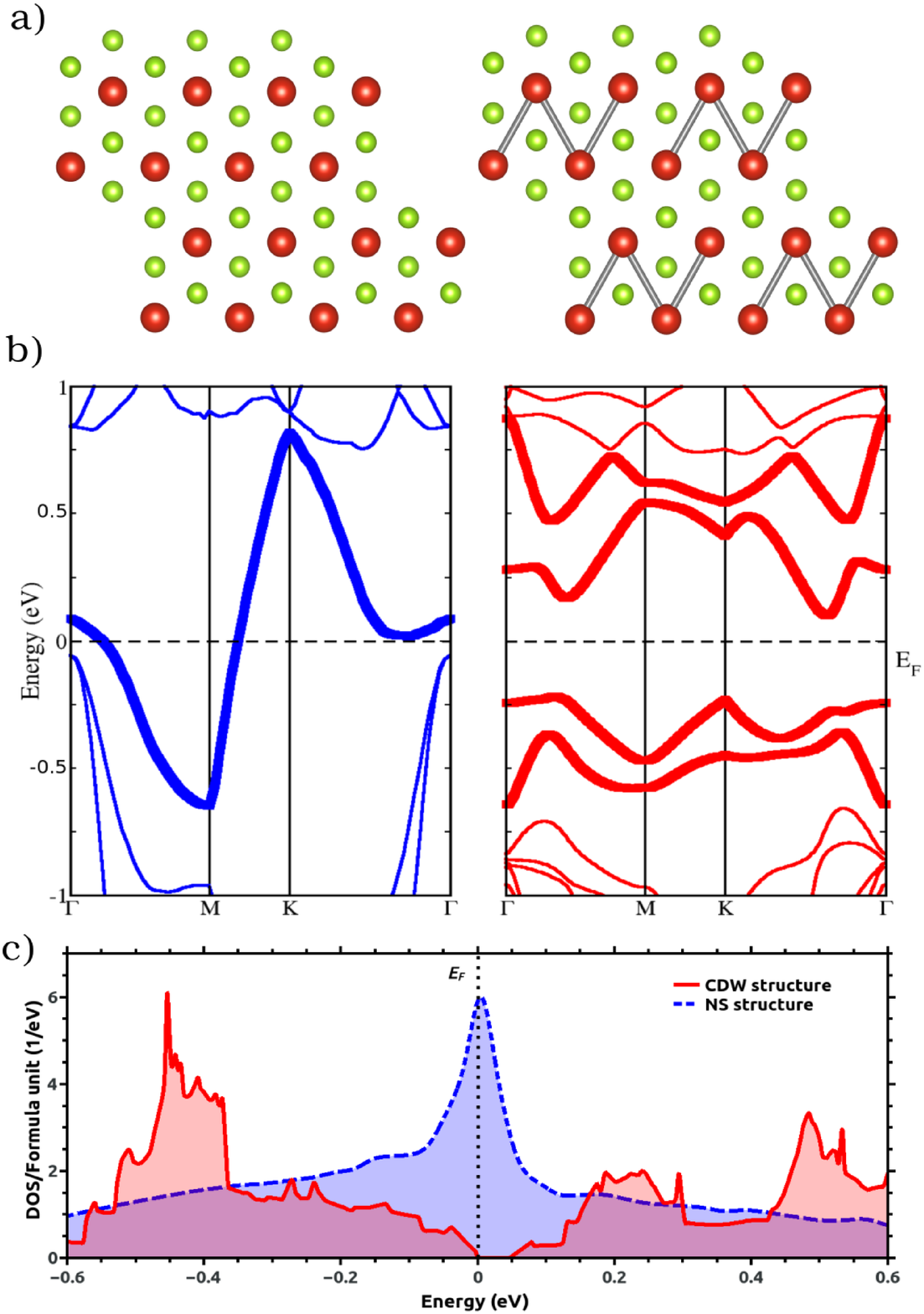}
\end{center}
\caption{(Color online.) Results for the monolayer structures. a) Monolayer structure schemes: NS structure in the left side, each V atom has 6 neighbor atoms at the same distance. 
CDW structure in the right side, a tetramer is formed in a 2$a$ $\times$ 2$a$ supercell. The tetramer bonds are depicted in grey. 
b) Monolayer band structures: In the left side the NS case, a d-band crosses the Fermi level producing a FM metallic state. 
In the right side the CDW case, the formation of the tetramer induces the creation of two bonding and two antibonding bands. As a result, an energy gap is open and FM is quenched. 
c) Monolayer DOS: the blue dashed line shows the NS DOS while the red line correspond to the CDW DOS. There is a clear gap opening in the CDW structure, causing a non-magnetic state.}
\label{mono_calc}
\end{figure}

Figure \ref{gapU} shows the evolution of the energy gap in the whole Brillouin zone as a function of U for a 2$a$ $\times$ 2$a$ supercell of the monolayer system. It can be seen that at low values of U a metallic FM NS structure is the most stable. 
For values of U greater than $2.7$ eV a non-magnetic CDW is formed and consequently an energy gap appears in the whole Brillouin zone at the Fermi level. 
In order to understand the features of this last solution, we have computed its band structure and DOS and compared it with the NS structure. 
The right side of Fig. \ref{mono_calc} a) shows the CDW structure scheme. 
In this solution a tetramer is formed between 4 V atoms as a consequence of a Peierls-like distortion, as has been experimentally determined\cite{nano_monolayer}. Its shorter V-V bonds are depicted in grey. 
This tetramer can be understood considering the band structures shown in Fig. \ref{mono_calc} b) and the DOS of panel c). 
In the left side the non-magnetic-monolayer NS band structure is shown. 
It can be seen that a d-band crosses the Fermi level. 
In the case that no structural distortion is allowed, this d-band and peaked DOS at the Fermi level (blue dashed line in panel c)) will lead to a FM state. 
However, when a CDW is considered, the 4 d-bands in the 2$a$ $\times$ 2$a$ supercell hybridize forming 2 bonding bands and 2 antibonding bands (right side of Fig. \ref{mono_calc} b)). 
This opens an energy gap (red line in panel c)) and quenches the FM moment.

A comparison between the DOS of the NS structures both for the bulk and the monolayer reveals that decreasing dimensionality increases the DOS at the Fermi level, the bands become flatter due to the absence of the off-plane hopping. In general, for itinerant systems, this would be a mechanism to enhance the stability of a ferromagnetic phase if the Stoner criterion can be applied to the system. However, this is not the case of VSe$_2$  because of the lattice distortion associated to the CDW state at low temperatures. This produces a gap opening at the Fermi level associated to a Peierls distortion with zero total magnetization of the system.

In conclusion, we have shown that DFT calculations predict that the ground state of bulk VSe$_2$ presents a commensurate lattice distortion with a 4$a$ $\times$ 4$a$ $\times$ 3$c$ supercell.
This structure is shown to be related with the CDW phase that has been experimentally detected at low temperatures. 
Our calculations for such a ground state show that ferromagnetism is destroyed by the distortion, VSe$_2$ being a paramagnet in the bulk. Such a structural change can also help understanding the change in thermopower observed experimentally at the transition.  
In the monolayer limit, a periodic lattice distortion (2$a$ $\times$ 2$a$) associated to the CDW state opens an energy gap through a Peierls like distortion destroying the tendency towards a FM state, that is found for the undistorted lattice.
Our calculations suggest that the origin of the magnetic signal obtained for VSe$_2$ cannot be intrinsic to the material, either in the bulk or in the single-layer limit.

The results that we have shown for VSe$_2$ could be extended to similar systems in which a CDW appears. 
Those demonstrate the importance of considering the correct ground state structure when performing \emph{ab initio} magnetic studies in this kind 
of compounds. 

This work is supported by the MINECO of Spain through the project MAT2016-80762-R. A.O.F. thanks MECD for the financial support received through the FPU grant FPU16/02572. We also thank S. Blanco-Canosa, J. Fernández-Rossier, I. Oleynik, Warren E. Pickett and D. Soriano for fruitful discussions. 


%

\end{document}